\font\twelverm = cmr10 scaled\magstep1 \font\tenrm = cmr10
       \font\sevenrm = cmr7
\font\twelvei = cmmi10 scaled\magstep1
       \font\teni = cmmi10 \font\seveni = cmmi7
\font\twelveit = cmti10 scaled\magstep1 
       
\font\twelvesy = cmsy10 scaled\magstep1
       \font\tensy = cmsy10 \font\sevensy = cmsy7
\font\twelvebf = cmbx10 scaled\magstep1 \font\tenbf = cmbx10
       \font\sevenbf = cmbx7
\font\twelvesl = cmsl10 scaled\magstep1
\font\twelvett = cmtt10 scaled\magstep1
\font\mbf = cmmib10 scaled\magstep1
       \font\mbfs = cmmib10 \font\mbfss = cmmib10 scaled 833
\font\msybf = cmbsy10 scaled\magstep1
       \font\msybfs = cmbsy10 \font\msybfss = cmbsy10 scaled 833

%
\textfont0 = \twelverm \scriptfont0 = \twelverm
       \scriptscriptfont0 = \tenrm
       \def\rm{\fam0 \twelverm}
\textfont1 = \twelvei \scriptfont1 = \twelvei
       \scriptscriptfont1 = \teni
       
\textfont2 = \twelvesy \scriptfont2 = \twelvesy
       \scriptscriptfont2 = \tensy
       
\newfam\itfam \def\it{\fam\itfam \twelveit} \textfont\itfam=\twelveit
\newfam\slfam  \textfont\slfam=\twelvesl
\newfam\bffam \def\bf{\fam\bffam \twelvebf} \textfont\bffam=\twelvebf
       \scriptfont\bffam=\twelvebf \scriptscriptfont\bffam=\tenbf
\newfam\ttfam  \textfont\ttfam=\twelvett
\textfont9 = \mbf
       \scriptfont9 = \mbfs \scriptscriptfont9 = \mbfss
       \def\bmit{\fam9 }
\textfont10 = \msybf
       \scriptfont10 = \msybfs \scriptscriptfont10 = \msybfss
       
\rm
\hsize=6.5in
\hoffset=.1in
\vsize=9in
\voffset=0in
\baselineskip=24pt
%
\raggedright  \pretolerance = 800  \tolerance = 1100
\raggedbottom
%
\dimen1=\baselineskip \multiply\dimen1 by 3 \divide\dimen1 by 4
\dimen2=\dimen1 \divide\dimen2 by 2
%
\def\apjsingle{\baselineskip = 14pt
               \parskip=14pt plus 1pt
               \dimen1=\baselineskip \multiply\dimen1 by 3 \divide\dimen1 by 4
               \dimen2=\dimen1 \divide\dimen2 by 2
               \scriptfont0 = \tenrm  \scriptscriptfont0 = \sevenrm
               \scriptfont1 = \teni  \scriptscriptfont1 = \seveni
               \scriptfont2 = \tensy \scriptscriptfont2 = \sevensy
               \scriptfont\bffam = \tenbf \scriptscriptfont\bffam = \sevenbf
               \rightskip=0pt  \spaceskip=0pt  \xspaceskip=0pt
               \pretolerance=200  \tolerance=400
              }
%
\nopagenumbers
\headline={\ifnum\pageno=1 \hss\thinspace\hss
     \else\hss\folio\hss \fi}
%
\def\heading#1{\vbox to \dimen1 {\vfill}
     \centerline{\bf #1}
     \global\count11=96  
     \vskip \dimen1}
%
\count10 = 0
\def\section#1{\vbox to \dimen1 {\vfill}
    \global\advance\count10 by 1
    \centerline{\expandafter{\number\count10}.\ \bf {#1}}
    \global\count11=96  
    \vskip \dimen1}
%
\def\subsection#1{\global\advance\count11 by 1
    \vskip \parskip  \vskip \dimen2
    \centerline{{\it {\char\number\count11}\/})\ {\it #1}}
    \vskip \dimen2}
%
%
\def\refindent{\advance\leftskip by 24pt \parindent=-24pt}
%
\def\journal#1#2#3#4#5{{\refindent
                      {#1}        
                      {#2},       
                      {#3\/}, 
                      {#4},   
                      {#5}.       
                      \par }}
%
\def\infuture#1#2#3#4{{\refindent
                  {#1}         
                  {#2},        
                  {#3\/},  
                  {#4}.        
                  \par }}
%
\def\inbook#1#2#3#4#5#6#7{{\refindent
                         {#1}         
                         {#2},        
                      in {\it #3\/},  
                     ed. {#4}         
                        ({#5}:        
                         {#6}),       
                       p.{#7}.        
                         \par }}
%

%
\def\book#1#2#3#4#5{{\refindent
                   {#1}         
                   {#2},        
                   {\it #3\/}   
                  ({#4}:        
                   {#5}).       
                   \par }}
%

%

%

%
\def\figcap#1#2{{\refindent
                   Fig. {#1}.---   
                        {#2}       
                        \par}}
%

%
\def\etal{{\it et al.\/\ }}
\def\eg{{\it e.g.,\/\ }}

\def\lsim{\raise0.3ex\hbox{$<$}\kern-0.75em{\lower0.65ex\hbox{$\sim$}}}
\def\gsim{\raise0.3ex\hbox{$>$}\kern-0.75em{\lower0.65ex\hbox{$\sim$}}}
\def\frac#1/#2{\leavevmode
	\kern.1em \raise .25em \hbox{\the\scriptfont0 #1}%
	\kern-.1em $/$%
	\kern -.15em \lower .125ex \hbox{\the\scriptfont0 #2}}%
\def\today{\ifcase\month
           \or January   \or February \or March    \or April
           \or May       \or June     \or July     \or August
           \or September \or October  \or November \or December\fi
           \space\number\day, \number\year}
\apjsingle
\centerline{\bf The Evolution and Efficiency of Oblique MHD Cosmic-Ray Shocks:}
\centerline{\bf Two-Fluid Simulations}
\vskip  1cm
\centerline{Adam Frank$^1$, T. W. Jones$^1$ and Dongsu Ryu$^{2}$}
\bigskip

\line{$^1$Department. of Astronomy, University of Minnesota, Minneapolis,

MN 55455\hfill}
\line{$^2$Department of Astronomy \& Space Science, Chungnam National
University,\hfill}
\line{\hskip 0.75cm Daejeon 305-764, Korea\hfill}
\bigskip
\bigskip
\bigskip
\bigskip

\heading{Abstract}

Using a new, second-order accurate numerical method we present dynamical
simulations of oblique MHD cosmic ray (CR) modified plane shock evolution using
the
two-fluid model for diffusive particle acceleration.  The numerical shocks
evolve to published analytical
steady state properties.  In order to probe the dynamical role of
magnetic fields we have explored for these time asymptotic
states the parameter space of upstream fast mode Mach number, $M_f$,
and plasma $\beta$, compiling the results into maps of
dynamical steady state CR acceleration efficiency, $\epsilon_c$.
These maps, along with additional numerical experiments, show that
$\epsilon_c$ is reduced through the action of compressive work on
tangential magnetic fields in CR-MHD shocks.  Thus $\epsilon_c$ in low
$\beta$, moderate $M_f$ shocks tends to be smaller in quasi
perpendicular shocks than it would be high $\beta$ shocks of the same
$M_f$.  This result supports earlier conclusions that strong, oblique
magnetic fields inhibit diffusive shock acceleration. For quasi
parallel shocks with $\beta < 1$, on the other hand, $\epsilon_c$
seems to be increased at a given $M_f$ when compared to high $\beta$
shocks.  The apparent contradiction to the first conclusion results,
however, from the fact that for small $\beta$ quasi parallel shocks,
the fast mode Mach number is not a good measure of compression through
the shock. That is better reflected in the sonic Mach number, which is greater.
Acceleration efficiencies for high and low $\beta$ having comparable
sonic Mach numbers are more similar.

Time evolution of CR-MHD shocks is qualitatively similar to
CR-gasdynamical shocks.  However, several potentially interesting
differences are apparent. We have run simulations using constant, and
non-isotropic, obliquity dependent forms of the diffusion coefficient
$\kappa$.  Comparison of the results show that while the final steady
states achieved are the same in each case, the history of CR-MHD
shocks can be strongly modified by variations in $\kappa$ and,
therefore, in the acceleration timescale.  Also, the coupling of CR and
MHD in low $\beta$, oblique shocks substantially influences the transient
density
spike that forms in strongly CR-modified shocks.  We find that
inside the density spike a MHD slow mode wave is generated that eventually
steepens into a shock.  A strong shear layer
develops within the density spike, driven by MHD stresses. We conjecture
that currents in the shear layer could, in non-planar flows,
result in enhanced particle acceleration through drift acceleration.

\vfill\eject

\section{Introduction}

In the past several years nonlinear theories of diffusive shock
acceleration have demonstrated the potential importance of cosmic-ray
(CR) dynamical feedback on the evolution of shock structures (see
reviews by Blandford \& Eichler 1987; Berezhko \& Krymskii 1988;
Jones \& Ellison 1991).  A range of methods has been successfully
applied to show that CR pressures can become sufficient to modify
shock structures substantially (\eg Drury \& V\"olk 1981; Achterberg
\etal 1984; Ellison \& Eichler 1984; Webb, Drury, and V\"olk 1986;
Falle \& Giddings 1987; Kang \& Jones 1990, 1991). In some cases the
modifications can even produce smooth CR dominated shocks with no
entropy generating gas sub-shock.  Numerical simulations using a variety
of techniques have
also demonstrated the importance of time-dependent effects in
the determination of CR modified shock properties (e.g., Drury \&
Falle 1987; Falle \& Giddings 1989; Kang \& Jones 1991; Duffy 1992;
Kang, Jones, \& Ryu, 1992; Ryu, Kang \& Jones 1993).

Through all of this the presence of a large-scale magnetic field was
generally understood to be an essential physical ingredient in CR
shocks, since magnetic field fluctuations (small-scale Alfv\'en waves,
or "Alfv\'enic turbulence" to avoid confusion with large scale waves)
mediate the CR propagation. But the magnetic fields are difficult to
treat explicitly and fully, especially when those fields play a direct
dynamical role.  So, most studies of CR modified shocks have focused
on pure gasdynamical models of the flows.  That may be adequate if the
fields are sufficiently weak, and/or the large-scale field is aligned
parallel to the shock normal, so that sonic-mode shock dynamics
applies. Except for modeling the spatial dependence of the CR
diffusion coefficient, explicit treatment of the magnetic field may
not always be generally necessary when the shock speed in the preshock
(upstream) fluid, $u_s$, and the preshock sound speed, $a$ are large
compared to the preshock Alfv\'en speed, $b$. The second of these
conditions is commonly expressed as $\beta = a^2/b^2 \gg 1$.  But,
this condition is likely to be violated, perhaps strongly, in many
environments where particle acceleration occurs (\eg Slavin \& Cox 1992).
Even if
the magnetic field and shock normal are parallel, treatment of CR
shocks in flows with $\beta \lsim 1$ may depend in significant ways on
$\beta$, especially if the Alfv\'enic Mach number is not very large.
One must be concerned, for example, with the finite momentum and
energy carried by the CR-mediating Alfv\'enic turbulence (\eg V\"olk,
Drury \& McKenzie 1984; Jones 1993a)

Postponing for now discussion of the more complicated topics in
general magnetohydrodynamical (MHD) flows, our purpose in the present paper is
to
examine  some straightforward dynamical influences of oblique magnetic
fields on plane CR shocks.  There are several obvious ways in which
oblique fields may be important.  If the pressure in the tangential
field is  comparable to the gas pressure of the
incoming flow, that changes the jump conditions to be satisfied
through the shock and must, therefore, influence the transfer of
energy and momentum to CR within the shock (\eg Webb 1983). The
anisotropic nature of MHD stresses in oblique shocks leads to the
existence of fast, intermediate and slow wave mode families, each with
distinct behaviors.  That can complicate the shock properties and lead
to the generation of distinctly MHD time dependent structures in a
shock evolving in response to CR acceleration. This increase in the
degrees of freedom at the shock also augments the modes through which
the shock structure may become dynamically unstable (\eg Chalov 1988;
Zank \etal
1990). The CR diffusion coefficient may be anisotropic relative to the
field, affecting the rates at which CR are accelerated.  If the
diffusion is described by standard kinetic theory models, the
acceleration rate at oblique shocks could be much faster than at
otherwise comparable parallel shocks (Jokipii 1987; Ostrowski 1988),
although the resulting time-asymptotic test-particle form for the
particle spectrum is apparently unchanged (Bell 1978), at least for
nonrelativistic shocks (Kirk \& Heavens 1989).  This change in
the acceleration rate should influence the dynamical properties of the
shock at intermediate times, before a dynamical equilibrium is
achieved, even if the field is weak and not directly involved
dynamically.  On the other hand, Baring, Ellison \& Jones (1993) have
shown that the efficiency of thermal particle injection into the CR
population can be decreased by perpendicular magnetic
field components. From these examples it is clear that as a step to
understand the fundamental nature of diffusive shock acceleration in
more realistic astrophysical settings, full time dependent MHD calculations
are needed.

Two-fluid models of CR transport along the lines developed by Drury \&
V\"olk (1981) and others are an efficient means to begin such
explorations. These models treat transport of the CR through an energy
conservation equation derived from the kinetic, diffusion-advection
equation (\eg Skilling 1975). The CR momentum distribution function
itself is not followed in detail, except possibly through {\it a priori}
models (\eg Jones \& Kang 1993; Duffy \etal 1993). Two-fluid methods are
meant to be used primarily for dynamical studies. They enable one to calculate
economically the dynamical features of flows within the constraints
imposed by the need to estimate {\it a priori} some closure parameters
for the CR (see Drury 1983; Kang \& Jones 1991; Kang 1993).
In particular one must model the adiabatic index for the CR and the
energy-weighted mean diffusion coefficient.  To the extent that the
solutions depend on those parameters (see, \eg Achterberg \etal 1984; Kang
\& Jones 1990) in ways that are not modelled, two-fluid models should not
be used for quantitative prediction. On the other hand, the dependence of
two-fluid solutions on these parameters for gasdynamic CR problems has been
reasonably well studied (\eg Achterberg \etal 1984; Kang \& Jones 1990)
and the basic validity of two-fluid dynamical solutions compared to
more complete, diffusion-advection solutions for the CR momentum distribution
demonstrated (\eg Falle \& Giddings 1987;
Kang \& Jones 1991; Ryu, Kang \& Jones 1993). Some recent calculations
even suggest that relatively simple models connecting the properties
of the particle distribution to the closure parameters might provide adequate
and
internally self-consistent procedures for eliminating most
limitations (Duffy \etal 1993).  In any case, our purposes here are
to explore the dynamical consequences of the magnetic field in these
problems.  Two-fluid methods would seem to be both adequate and
appropriate for that aim.

Webb (1983) first applied
two-fluid methods to equilibrium, steady state CR-MHD modified shock
structures for the case where the upstream gas is cold (its
pressure can be ignored) and the magnetic field is oblique.  Webb's
analytic calculations demonstrated that, as
for the CR modified gasdynamical flows, both sub-shock containing and
smoothed, CR dominated solutions to the CR-MHD shock equations were
possible.  However, his solutions suggested that shock CR acceleration
was less effective when the upstream tangential component of the field
was strong. Subsequently Kennel, Edmiston \& Hada (1985) and Webb,
Drury, and V\"olk 1986 (hereafter, WDV) calculated steady state
two-fluid CR-MHD modified shock structures with finite upstream gas
pressures.  From their analysis, which included $\beta = 5$ and $\beta = 1$
cases and a range of fast mode Mach numbers, $M_f$, they concluded that
the efficiency of CR acceleration
was suppressed when the obliquity of the upstream magnetic field was
high.  They also found that the reduction in CR
acceleration efficiency was most pronounced in the lower $\beta$ ($\beta = 1$)
flows of low
fast mode Mach number ($M_f < 6$). For high $\beta$ ($\beta =
5$) flows the CR acceleration efficiency was only weakly affected by
changes in the upstream magnetic field angle regardless of the fast
mode Mach number.

As an extension of the previous work, we report in this paper the
results of {\it time dependent} CR-MHD two-fluid simulations using a
new numerical code.  In an earlier work (Frank, Jones, \& Ryu 1993) we
presented tests of this code against Webb's cold gas analytical models
as well as more general internal checks on our code's ability to
evolve oblique shocks to self-consistent steady state CR-MHD structures. The
simulations in that paper also confirmed Webb's cold gas solutions,
presented preliminary results from more general CR-MHD flows and
determined the convergence properties of our numerical method. In the
present paper we explore the time evolution of finite gas pressure
($P_g \ne 0$)
models of CR-MHD modified oblique shocks and their relation to the WDV
steady state calculations. We will also explore in more detail
questions raised in WDV concerning the role of both the obliquity of
the field and the tangential field strength in determining the CR
acceleration efficiency. More general time dependent effects unique to
CR-MHD modified shocks will also be examined along with their
potential importance for CR studies. As in WDV we will concentrate on
those flows with high enough Alfv\'enic Mach number that switch-on and
switch-off shocks do not occur.  In the regime where those special
shocks form, CR scattering centers (the Alfv\'enic turbulence) move at
speeds comparable to the gas, requiring that the time dependent wave
field dynamics be included in the calculations. For the present
we restrict ourselves to flows where the scattering centers can be
considered moving with the fluid.  Jun, Clarke \& Norman (1993)
have also recently simulated CR-MHD shocks using two-fluid methods. Their
results are complementary to ours, since they restricted their
analysis to the special case of perpendicular (magnetosonic) shocks, so
that they could not explore a number of questions address here.

The remainder of the paper is organized as follows.  Section 2
introduces the fundamental equations governing CR modified MHD flows
and discusses their application to our methods. An appendix at the end
discusses these numerical methods in more detail.  Section 3 compares
the results of our models with those of WDV and explores issues
concerning the dependence of CR shock acceleration on upstream
magnetic field properties.  In \S 4 we examine the general time
dependent features of CR-MHD modified shocks as seen in our
simulations, while we provide in \S 5 a discussion of these results in
light of previous studies and their potential for application in
higher dimensional flows.  We present our conclusions in \S 6.

\section{Methods}

We solve the equations of non-relativistic ideal MHD for one
dimensional flow in Cartesian coordinates (\eg Jeffrey 1968).  As with
gasdynamical models the conservation equations are modified
to include momentum and energy source terms from CR feedback (\eg
Drury \& V\"olk, 1981; Jones \& Kang, 1990).  The MHD equations are
written in conservative, vector form as $${\partial{\bmit
q}\over\partial t}+{\partial{\bmit F}\over
\partial x}={\bmit S}, \eqno(2.1)$$
with
{\textfont1 = \twelvei
       \scriptfont1 = \twelvei \scriptscriptfont1 = \teni
       
$${\bmit q} = \left(\matrix{\rho\cr
                          \rho u_x\cr
                          \rho u_y\cr
                          \rho u_z\cr
                          B_y\cr
                          B_z\cr
                          E\cr}\right), \eqno(2.2)$$}
and
{\textfont1 = \twelvei
       \scriptfont1 = \twelvei \scriptscriptfont1 = \teni
       
$${\bmit F} = \left(\matrix{\rho u_x\cr
                          \rho u_x^2+P^*-B_x^2\cr
                          \rho u_x u_y-B_x B_y\cr
                          \rho u_x u_z-B_x B_z\cr
                          B_y u_x-B_x u_y\cr
                          B_z u_x-B_x u_z\cr
            (E+P^*)u_x-B_x(B_x u_x+B_y u_y+B_z u_z)}\right),
\eqno(2.3)$$}
while the CR source term vector is
{\textfont1 = \twelvei
       \scriptfont1 = \twelvei \scriptscriptfont1 = \teni
       
$${\bmit S} = -  \left(\matrix{0\cr
                          \partial P_c/\partial x\cr
                          0\cr
                          0\cr
                          0\cr
                          0\cr
                          u_x\partial P_c/\partial x + S_e\cr}\right).
\eqno(2.4)$$}
In two-fluid models the CR are themselves treated as a massless, diffusive
fluid through a
conservation equation for the CR energy, $E_c$, derived by taking the
second moment of the diffusion-advection equation (\eg Achterberg
1982); namely,
$${{\partial E_c}\over{\partial t}}~+~{{\partial}\over
{\partial x}}\left ( u_x E_c~- ~\langle \kappa\rangle{{\partial E_c}\over
{\partial
x}}\right ) = -~P_c {{\partial u_x}\over {\partial x}}~+~
S_e.\eqno(2.5)$$
A more general treatment would replace equation [2.5] with the
diffusion-advection
equation itself.  Its moments would be calculated independently.
In these relations
$$P^* = P_g + {1 \over 2} (B^2_x +
B^2_y + B^2_z) = P_g + {1 \over 2} B^2,\eqno(2.6)$$
$$E = {1 \over 2}
\rho(u^2_x+u^2_y+u^2_z)+ {1 \over {(\gamma_g-1)}} P + {1 \over 2}
(B^2_x+B^2_y+B^2_z),\eqno(2.7)$$
$$P_c =(\gamma_c - 1) E_c,\eqno(2.8)$$
and the magnetic field components are expressed in
rationalized units
$$B \rightarrow {B \over \sqrt {4\pi}},\eqno(2.9)$$
so that the Alfv\'en speed, $b = B/\sqrt{\rho}$. In the expressions
presented above the following definitions hold: $\rho$ is the mass
density; $u_x$, $B_x$ and $u_y$, $u_z$, $B_y$, $B_z$ are the
components of velocity and magnetic field parallel and perpendicular
to shock front; $P_g$, $\gamma_g$ and $P_c$, $\gamma_c$ are the gas
and CR pressures and adiabatic indices, and $\langle\kappa\rangle$ is the
energy-weighted, CR diffusion coefficient parallel to
the shock normal (Drury 1983). (For simplicity, we will henceforth omit the
brackets around $\kappa$.)
It is not necessary in two-fluid models to assume that either of
$\gamma_c$ or $\kappa$ is a constant or that $\kappa$ derives from an
energy independent diffusion process. As described earlier, they
do depend on {\it a priori} models, however. The quantity $S_e$ is a term that
allows
direct energy transfer from the gas to the CR, such as through the
injection of thermal particles into the CR population (\eg Jones \&
Kang 1990).  This is introduced for completeness, but for the present,
we set $S_e = 0$. We note in passing, however, that in the course of
this investigation we carried out analogous calculations with $S_e \not = 0$.
Behaviors of these shocks were not qualitatively different from
those we will describe ({\it cf.} Kang \& Jones 1990, 1991).
In this discussion $\gamma_g$ = \frac{5}/{3}, while the CR
closure parameter $\gamma_c$, which depends upon the mix of
non-relativistic ($\gamma_c \rightarrow \frac{5}/{3}$) and relativistic
($\gamma_c \rightarrow \frac{4}/{3}$) particles in the CR population, will be
treated as an input parameter.  In general both $\gamma_c$ and
$\kappa$ are properties of the solution, so the need to specify their
values {\it a priori} is the principal restriction of the two-fluid
model.

In the interest of simplicity we will not consider here flows with
rotations of the shock-normal/magnetic-field plane between the initial
upstream and downstream states, although the numerical code is quite
capable of handling such features. Thus, without further loss of
generality we can place the magnetic field in the X-Z plane, $\vec B =
(B_x,0,B_z) = B(cos{\theta}, 0, sin{\theta})$.  We will also select a
reference frame so that there is
no upstream tangential velocity; $u_y = u_z = 0$.  All of the
simulations discussed here are piston driven shock tubes.  We
establish the flows by projecting magnetized fluid with embedded CR in
from the right boundary, using continuous boundary conditions and
reflecting it off a wall (piston) at the left boundary. The tangential
magnetic field at the left boundary is ``mirrored'' in a manner
analogous to the gas density. Thus, there is no current sheet on the
piston surface.
In other words, the simulations can be considered as those of two
identical, colliding clouds with magnetic field ``continuous'' across
them. The CR pressure is given continuous boundary conditions on both
ends of the grid.

In general, spatial diffusion will differ in directions parallel and
perpendicular to the magnetic field.  Diffusive acceleration depends
on the diffusion along the shock normal, since that determines the
rate at which particles re-cross the shock.  Thus, the appropriate
diffusion coefficient in equation [2.5] will take the form (\eg Drury
1983; Webb 1983),
$$\kappa = {\kappa_\parallel} \cos^2 \theta + %
{\kappa_\perp} \sin^2 \theta, \eqno(2.10)$$
where the directions
$\parallel$ and $\perp$ refer to the magnetic field direction.  Note,
of course, that $\theta$ is generally a changing function of space and
time.  In standard kinetic theory scattering models of particle
diffusion the ratio $\kappa_\parallel / \kappa_\perp \sim 1~+~ (\omega
\tau)^2$, where $\omega$, and $\tau$ are the particle gyro frequency
and collision time, respectively (\eg Jokipii 1987; Zank \etal 1990).
We will adopt the form for $\kappa$ in equation [2.10] in our
simulations.

Previous studies of CR modified shocks have shown that the time
required to evolve towards dynamical equilibrium scales with the
so-called diffusion time, $t_d$, (\eg Drury \& Falle 1986; Jones \&
Kang 1990), which in the present case is conveniently expressed as
$$t_d = {\kappa \over u^2_s} = {x_d \over u_s},\eqno(2.11)$$
where $u_s$ is the shock
speed in the upstream fluid (see equation [3.7]) and $x_d$ is known as
the diffusion length.  $x_d$ is a measure of the breadth of the CR precursor
to the shock.
It is important that computational resolution be fine enough to resolve the
flow features on this scale.  For our discussion we define,
therefore, the resolution ratio of
each simulation to be $$n_r = {x_d\over \Delta x },\eqno(2.12)$$
where $\Delta x$ is the size of a computational zone.
Frank, Jones \& Ryu (1993) found that provided $n_r~>~10-20$, our code
produced accurate and converged solutions. When $\kappa_{\parallel} \not =
\kappa_{\perp}$,
$t_d$ is not well defined, since $\kappa$ changes through a shock.
To be specific, however, we will adopt the convention of using the
value of $\kappa$ from equation [2.10] upstream of the shock to define
$t_d$ in our discussions.

Our numerical method solves equation [2.1] through the
second order accurate, Eulerian finite difference,
"Total Variation Diminishing" (TVD) scheme.
The TVD scheme was originally designed for gasdynamics by Harten (1983)
and extended to MHD by Brio and Wu (1988).
In this paper, we use the version of the MHD-TVD code which was
subsequently improved and fully tested by Ryu \& Jones (1993).
The pure MHD $({\bmit S} = 0)$ form of the equation is solved with
the aid of an approximate MHD Riemann solver used to estimate the
flux, ${\bmit F}$. CR source corrections, ${\bmit S}$, are added separately
in a manner preserving second order accuracy.  Shocks and other
discontinuities are generally sharply resolved within a few zones.
The CR energy equation is solved separately using a second order
combined Lagrangian Crank-Nicholson and monotone remap scheme. This
mixed scheme is efficient and provided us much better
stability and accuracy than a
straight Eulerian Crank-Nicholson approach.  In the appendix we
present the numerical method in more detail. A pure gasdynamical
version of the code was tested extensively against both analytical
steady state solutions and numerical time dependent models calculated
with our well-tested PPM two-fluid code (Jones \& Kang 1990), all
with excellent agreement. The pure MHD code was tested using a
nonlinear MHD Riemann solver against a variety of planar shock tube
problems involving all three wave mode families (for details, see
Ryu \& Jones 1993).

\section{Results}

We will now summarize results from these simulations, beginning with
some properties of flows that were continued until the state immediately
downstream of the shock was no longer evolving; {\it i.e.,} a time
asymptotic state.  Then we will
explore behaviors of CR-MHD shocks at intermediate times, while
the post shock flow is still developing. There are also some notable
features that continue to evolve downstream of the equilibrium
shock.

To describe the flows there are four parameters needed in addition
to the CR "fluid" properties, $\gamma_c$ and $\kappa$.  Thus,
we parameterize our upstream state by the following dimensionless
quantities,
$$\beta = {a^2 \over b^2} = {{\gamma_g P_g} \over{ B^2}}$$
$$N = {P_c \over {P_g + P_c}},\eqno(3.1)$$ $$M_f = {\vert u_x \vert
\over c_f},$$
$$\theta_o = \tan^{-1}({{B_z}\over{ B_x}})$$.

In equation [3.1] all quantities refer to the upstream (incoming flow)
state and $u_x$ is measured in the computed (piston) frame.  In the following
discussion wherever it is necessary to distinguish upstream and downstream
conditions they will receive subscripts $_1$ and $_2$ respectively.
Of the remaining variables in equation [3.1]
$a$ is the gas sound (acoustic) speed, $c_f$ is the fast mode wave speed;
namely,
$$a = \sqrt{{\gamma_g P_g \over \rho}},\eqno(3.2)$$
$$c^2_f = {1 \over 2}\lbrace b^2 + a^2 + [(b^2 +
a^2)^2 - 4b^2a^2\cos^2 \theta_o] ^{{1 \over 2}}\rbrace ,\eqno(3.3)$$
and $\theta_o$ is termed the "obliquity" of the shock.
Note that the fast mode Mach number, $M_f$, in our simulations refers to
the piston rather than the shock. The equivalent shock Mach number is
$$M_{fs} = {\vert u_s \vert \over \vert u_x \vert }M_f.\eqno(3.4)$$
The relationship between $M_f$ and $M_{fs}$ depends on the compression
through the shock, which is generally a function of the solution itself
and a function of time as $P_{c2}$ changes. For most of the shocks
described below, $M_f < M_{fs} \lsim \frac{4}/{3} M_f$.
Our definition of $\beta$ differs from that used frequently in the plasma
physics literature and by WDV by a factor \frac{$\gamma_g$}/{2} = \frac{5}/{6}.
For all the simulations described, the upstream flow is
characterized by $\rho = 1$, $u_x = -1$ and $N = \frac{1}/{2}$.

\subsection{Steady CR-MHD Shocks: Comparison with Previous Calculations}

We start with a comparison between properties of a steady state model
computed by WDV and the approximately equivalent time asymptotic flow found by
us.
Previously (Frank, Jones, \& Ryu 1993) we showed good agreement between our
results and the special
``cold gas'' analytic solutions examined by Webb.
These comparisons are useful not only as tests of numerical methods,
but also as demonstrations that the analytic steady states can be
reached from some other initial state. In that context, however, we do not
address in
this paper
issues associated with multiple steady state solutions found by WDV for
CR-MHD shocks (see Donohue \etal 1993 for such a discussion regarding
CR-gasdynamic shocks).

Figure 1 illustrates the evolution to an apparent dynamical steady post shock
state of a simulated flow with upstream parameters
$\beta = 1$, $\gamma_c = \frac{4}/{3}$, $\kappa = 10$, and
$\theta_o = 60^o$. The resolution ratio was
$n_r = 33$, so that the solution is well converged.
The simulation was designed to match as closely as possible a
solution presented by WDV in which $M_{fs} = 6.1$. (Note that their
definition of $c_f$ included $P_c$ as a contributor to the sound speed.
We have corrected for that in establishing an appropriate comparison.)
The downstream partial pressures, $P_{c2}$, $P_{g2}$ and $P_{B2}$,
from WDV are indicated by dashed lines.
The final $u_s$ (and thus $M_{fs}$) is hard to predict exactly,
because it depends on $P_{c2}$, $P_{g2}$ and $P_{B2}$ in nonlinear ways.
In addition the WDV solution was read from a published figure (Figure 12
in DWV), so we cannot
match the two problems exactly.  Nonetheless we find that all the partial
pressures agree to within about 1\% when normalized by the total momentum
flux incident on the shock, ${\bar P}$, defined as
$${\bar P} = \rho_1 u^2_s + P_{B1} + P_{g1} + P_{c1},\eqno(3.5)$$
where to estimate $u_s$ in the final steady shock, we
assume the mass flux jump conditions are exactly
satisfied across the shock.  Then the required $u_s$ in the frame of the
upstream fluid
can be written in terms of jump conditions across the shock in the computed
frame as
$$u_s = {[\rho u_x] \over [\rho]} - u_{x1} = {r\over {r~-~1}}\vert
u_{x1}\vert,\eqno(3.6)$$
where
$$r = {\rho_2 \over \rho_1},\eqno(3.7)$$
is the compression ratio through the full shock structure.

Figure 1 shows the flow at times $t/t_d =
325,~856$ and $1377$.  The shock transition has approached
a steady state structure by $t/t_d = 856 $.  The rather large number of
diffusion times needed for this shock to come into a dynamic
equilibrium is consistent with previous gasdynamical CR shock
calculations (\eg Jones \& Kang 1990; Kang \& Jones 1991). Jones \&
Kang (1990) pointed out that the rate for CR pressure to increase in
response to diffusive acceleration depends not only on $t_d$, but also
inversely on the CR adiabatic index. Physically that derives from the fact
that pressure generated by non-relativistic particles scales as the
particle momentum squared, while it increases only linearly with
momentum for relativistic particles.  Since $\gamma_c = \frac{4}/{3}$, the
pressure in this model is assumed to come entirely from relativistic
particles.  The time to reach equilibrium also depends on the total
change in $P_c$, of course, and so on $N$ and $M_{fs}$.

As $P_c$ builds within the shock precursor, a postshock density spike
forms and then is left behind after the shock approaches equilibrium.
The origin of this feature in strongly modified gasdynamic CR shocks was
explained by
Drury (1987) and independently by Jones \& Kang (1990).  Briefly, it
reflects an adiabatic compression in the precursor at times
before an initially strong subshock weakens in response to substantial
adiabatic heating in the
precursor. The tangential magnetic field participates in the
development of that feature in oblique MHD shocks, as can be seen in Figure 1.
These features are examples of time dependent effects that occur in
the evolution of CR modified MHD shock structures and of which the
steady analytical models can give no hint.  We will discuss these
time-dependent features in more detail in section 3e.

\subsection{Steady CR-MHD Shocks: CR Acceleration Efficiency Maps}

We now explore the influence of tangential magnetic fields on the CR
acceleration efficiency.  All previous related discussions have pointed out
that
strong, oblique fields tend to inhibit acceleration efficiency, at least
in shocks of modest strength.  However, the physical reasons for this
have not been well determined. Based on their study of magnetosonic
shocks Jun \etal (1993) made the reasonable conjecture that efficiency was
reduced because a tangential field inhibits compression by stiffening
the equation of state for the fluid .  But the physics of oblique
CR-MHD shocks is complex and nonlinear, so this question needs further
examination to be understood.  For this discussion we
will define the CR acceleration efficiency to be the ratio of
downstream CR pressure to upstream total momentum flux,
$$\epsilon_c = {P_{c2} \over \bar P}.\eqno(3.8)$$
In their
study, WDV used analytical solutions to the steady CR-MHD shock equations
for $M_f = 5$ and $1$ and $\theta_o = 0^o, \ 30^o$ and $60^o$ to conclude
that the obliquity of the field was the important acceleration inhibiting
factor.
To extend that effort we have completed a large set of simulations designed
to map out time
asymptotic values of $\epsilon_c$ as a function of the upstream
parameters.  A compact, visual way to examine how $\epsilon_c$
depends on the magnetic field properties in the flow is to construct maps
of $\epsilon_c$ on a plane defined by
the in-flowing normal and tangential Alfv\'enic
Mach numbers,
$$M_{x} = {u_x \over b_{x}} = {u_x \over b \cos
\theta_o} = {M_g \beta^{1\over 2} \over \cos \theta_o}
\propto {1 \over B_x}, \eqno(3.8a)$$
and
$$M_{z} = {u_x \over b_{z}} = {u_x \over b \sin \theta_o} =
{M_g \beta^{1\over 2} \over \sin \theta_o}
\propto {1 \over B_z}, \eqno(3.8b)$$
where $M_g$ is the sonic (gas) Mach number.  For each map
$\beta$ is held constant. Shocks with
a range of shock fast mode Mach numbers, $M_{fs}$, are
obtained by varying $M_{x}$ and $M_{z}$ .  For this purpose it is
convenient to write $M_f$ in terms of $\beta$ and the total
Alfv\'enic Mach number, $M_b$,
$$M_f^2 ={{2 M_b^2}\over { \lbrace 1 +
\beta + [(1 + \beta)^2 - 4\beta\cos^2 \theta_o]
^{1 \over 2}\rbrace}} ,\eqno(3.9)$$
where
$${1 \over M_b^2} = {1 \over M_{x}^2} +
{1 \over M_{z}^2} = {{b^2}\over{u^2_x}} = {1 \over{ \beta
M^2_g}}.\eqno(3.10)$$
It is also useful to note that
$$M_b = M \sin 2\theta_o,\eqno(3.11a)$$
$$\theta_o = tan^{-1}(M_x/M_z), \eqno(3.11b)$$
and
$$M^2_z = {\rho u_x^2 \over P_B}~~ _{\overrightarrow{N=1/2}}~~ {\bar P \over
P_B} - 1 - 2\beta, \eqno(3.11c)$$
where
$$M = (M_{x}^{2} + M_{z}^{2})^{1\over 2}.\eqno(3.11d)$$

Figure 2 presents such time-asymptotic efficiency maps for $\beta =
20,~3,~0.8,$ and $0.2$.  Each map was made from 25 simulations which uniformly
covered the ($M_x,M_z$) plane to include $2 \le M_{fs} \le 16$. All of the
simulations have
$\gamma_c = 1.41$ and $n_r > 20$.  The contours of CR acceleration
efficiency, $\epsilon_c$, are shown as solid lines and are labeled.  The
contours of fast mode shock Mach number, $M_{fs}$, are shown as dotted lines.
They begin with the lowest contour set at $M_{fs} = 2$ in the lower left
corner, increasing in steps of 2.
Although the shocks in $\beta < 1$, $M_{fs} \sim 2$ flows would likely be
influenced
by finite ``Alfv\'en transport'' effects (\eg Jones 1993a), their inclusion
here is useful in establishing trends within the CR-MHD solutions.

Some preliminary explanation should make the physics illustrated in
Figure 2 easier to grasp.  Note first that moving along an arc of
constant radius ($M$) from the vertical axis corresponds to sweeping
through $\theta_o$ from $0^o$ to $90^o$. The total Alfv\'enic Mach number,
$M_b$, and the fast mode Mach number, $M_f$, are maximum on this arc at
$\theta_o = 45^o$.  We note, as well, that
for large $M_z$ contours of $M_f$ asymptote to lines $M_x = M_f$, while
for large $M_x$ these contours asymptote to lines
$M_z = \sqrt{1 + \beta} M_f$.
Thus parallel and perpendicular shocks lie only at infinity in
this plane.

It is easiest to begin by examining the $\beta = 20$ map, since MHD effects
are minimal.  There one can see that both
the $\epsilon_c$ and $M_{fs}$ contours are almost symmetric about
$\theta_o = 45^o$ and that they are almost ``parallel''.  There is a
weak tendency at large obliquity for $\epsilon_c$ to decrease along
a Mach number contour as $\theta_o$ increases.  This is consistent with
the conclusion reached by WDV.  The effect is seen to be much stronger
as one looks at successively smaller $\beta$; {\it i.e.,} as the field
becomes dynamically more important.
Consider, for example, the obliquity at which $\epsilon_c$ drops
below 0.6 on the $M_{fs} = 8$ contour.  As $\beta$ decreases
from 20 to 0.2 the value of the upstream obliquity at which these
contours cross drops from $\theta_o \approx 63^o$ to $\theta_o \approx 47^o$.
Thus, as the magnetic field becomes more dynamically significant shocks
of constant $M_{fs}$ and constant $\theta_o$ become less efficient CR
accelerators found by others.  Relation [3.11c] provides a clue to what is
happening.
Lines of constant $M_z$ correspond to a constant ratio of in-flowing
momentum flux to tangential magnetic pressure.  When $\beta$ is small
$P_B/\bar P \propto M^{-2}_z$. For $\beta < 1$, $\epsilon_c$ contours
in Figure 2 for high obliquity (quasi perpendicular) shocks are closely
spaced and roughly parallel to the $M_x$ axis, and so parallel to
constant $P_B/\bar P$. Thus, it is the dynamical capacity of
magnetic pressure that seems
to be most significant. This is reasonable and consistent with previous
conclusions, of course.  The effect is important only for small to moderate
$M_{fs}$, since as $M_{fs}$ becomes large $M_z$ is
restricted to values $> \sqrt{1+\beta} M_{f} \gg 1$. Then from equation [3.11c]
$P_B/\bar P$ must be small for any $\beta$. This conclusion still
does not discriminate among at least two possible physical reasons
why the magnetic pressure plays this role.  It could be, as suggested
by Jun \etal (1993) just that the magnetic field reduces compression
through the shock.  Alternatively, it could reflect the need for
the shock to do work on the field in compressing it, thus reducing the
energy available to CR.

We will describe an experiment in the following subsection that may
help to resolve that issue.  But, first we should comment on another
aspect of the efficiency contours in Figure 2. In this case we focus on
quasi parallel shocks, so that $M_x$ is relatively small.  The efficiency
of these shocks in the strong field ($\beta < 1$) regime seems to be
higher at a given fast mode Mach number than when the field is weak.
For instance, when $\theta_o = 20^o$ and $\beta = 0.2$, a $M_{fs} = 4$
shock leads to $\epsilon_c \approx 0.55$.  At the same angle this efficiency
requires $M_{fs} > 6$ when $\beta = 20$. On the face of it this seems
to imply that the strong field has actually enhanced $\epsilon_c$. But
that would be a misinterpretation of the results.  When $\beta \ll 1$
$M_f \approx M_b$; {\it i.e.,} the fast mode wave is much like an
Alfv\'en wave for quasi parallel propagation and not very compressive.
The large value of $B$ that makes $c_f \approx b$ so large is mostly in the
normal component.  Using $M_z$ as an indicator again, we can see that the
dynamical impact of the tangential field cannot be very large in this regime.
Thus, the
fast mode Mach number underestimates the strength of the shock as it relates
to diffusive acceleration.
Indeed, examination of the structures of the CR-MHD shocks here
reveals that the time asymptotic compression is large ($r > 4$) and
comparable to
that found in large $\beta$ shocks having similar {\it acoustic} Mach
number $M_g = M_b/\sqrt{\beta}$.

This last feature should emphasize that it would be wrong to expect
that the existence of strong fields {\it necessarily} leads to a
reduction in the efficiency of CR acceleration in astrophysical
environments.  That statement seems
to apply only when the {\it tangential} magnetic pressure is strong. Thus,
the orientation of the field is particularly important as $\beta < 1$.

\subsection{Steady Solutions: The Role of Compression and Magnetic Pressure}

Diffusive shock acceleration results from simultaneous compression and
diffusion of CRs.  Thus, the compression through the shock is itself
obviously important to the efficiency of the process.  But nonlinear effects
complicate this simple statement. Even in CR-gasdynamical shocks the
acceleration efficiency represents a balance in the competition between
energy input into the CR and the gas.  The incoming momentum flux
does work on both the gas and the CRs, with the partitioning dependent
on the required entropy production through the structure.  In general
a simple relationship between total compression and efficiency does
not exist. When magnetic pressure is included the situation is even
more complex, since compression of the magnetic field also requires
work.

To explore the relative roles of the compression itself and the work done
on the magnetic field in CR-MHD shocks, we have performed
a simple set of numerical experiments. We carried simulations out
to dynamical steady state postshock conditions for two values of
the upstream obliquity, $\theta_o = 30^o$ and $60^o$ and two
values of $\beta = 5,~1$.  For each of these combinations
we computed solutions
for a range of CR adiabatic index, $\frac{4}/{3} \le \gamma_c \le
\frac{5}/{3}$.
By varying $\gamma_c$ we change the compression through the shock,
since $\gamma_c < \frac{5}/{3}$ softens the equation of state in a
manner analogous to the way that the magnetic pressure stiffens it
compared to the gas alone. All of the models used $M_f = 6$.

The results
are illustrated in Figure 3.  Figure 3a (upper left) shows the compression
ratio, $r$, for the $\theta_o = 60^o$ models as a function of $\gamma_c$.
As we expect from previous studies (\eg Drury \& V\"olk 1981; Achterberg
\etal 1984; Kang \& Jones 1990)
smaller $\gamma_c$ leads to substantially greater compression through
the shock.  Note, however, that the compression is independent of $\beta$
within our levels of uncertainty. As seen in Figure 3b (upper right)
the efficiency,
$\epsilon_c$, does depend on $\gamma_c$, and thus on the compression.
Note, however, that $\epsilon_c$  actually decreases with increasing
compression; {\it i.e.,}
as the mixed fluid develops a softer equation of state.  That trend
runs counter to the compression argument given by Jun \etal (1993) regarding
the role of the magnetic field.
But, $\epsilon_c$ {\it also} depends on $\beta$. More to the point,
for a fixed compression ($\gamma_c$), $\epsilon_c$ is reduced  as
the tangential magnetic field becomes stronger.  The change in $\epsilon_c$
can be accounted for roughly in the work done on the field itself.  That can
be seen in Figure 3d, which shows the normalized downstream magnetic
pressure, $P_{B2}/\bar P$, for the same study.  Two points are
relevant.  First, $P_{B2}/\bar P$ varies in a sense opposite to $\epsilon_c$
as $\gamma_c$ varies.  Second, the difference in $P_{B2}/\bar P$
between the weak and strong field cases is comparable to
that in $\epsilon_c$ for the same two cases.  Thus, crudely, at
least, there has been an exchange between these two quantities.
But, the nonlinear nature of these processes would lead us to expect
a more complex situation quantitatively. Indeed, for
the smaller obliquity case, $\theta_o = 30^o$, seen in Figure 3c,
the efficiencies for different $\beta$ are the same to within our $\sim 1$\%
uncertainties.  The magnetic pressure, $P_{B2}/\bar P$ is smaller and has
the same trend as in the other example.  But, the low and high $\beta$
values of $P_{B2}/\bar P$ differ by as much as 3\%. Clearly, the other
energy channels (gas pressure and kinetic energy) are also involved.

\subsection{Time-dependent Effects: Nonisotropic Diffusion }

{}From this point we will concentrate on evolutionary aspects
of CR-MHD shocks as we can understand them through two-fluid models.
For a variety of reasons it is likely that most astrophysically
important CR shocks will not be of a steady character (see, \eg
Jones 1992).

It is well known that the time asymptotic efficiency of two-fluid
CR-gasdynamical  shocks should be insensitive to the value or form of
$\kappa$ (Drury \& V\"olk 1981; Jones 1993b). On the other hand the detailed
properties of $\kappa$ do influence the rate of shock evolution and the
structure of the shock. We now briefly examine those issues for oblique
CR-MHD shocks.  Figure 4 presents properties of two simulations differing
only in the properties of $\kappa$.  The other properties of the flows were:
$\beta = 1$, $M_f = 6$, $\gamma_c = 1.41$, $\theta_o = 60^o$.
Both shocks are shown at the
same dynamical time, $t$, after the shock itself has reached an equilibrium
structure. The dotted line represents a shock
computed with constant $\kappa = 0.325$, while the solid line presents
a flow with an anisotropic (obliquity dependent) $\kappa$ with
$\kappa_{\parallel} = 1.0 = 10 \times \kappa_{\perp}$, as
defined in equation [2.10]. Upstream of the shock $\kappa = 0.325$ in
this model also.  Two points are obvious from the figure.  First,
the final post shock states, as represented by $\rho$ and $P_c$, are the same
in both simulations.  This holds for the other state variables, as well.
Second, the histories of the two shocks were
different.  One can trace the history by looking from left
to right, since the shock moves away from the piston at the left
boundary and to the first approximation the post shock fluid is
at rest and not changing very fast (the next section discusses ways
in which this is not quite true). The density spike identifies the
approximate location of the shock when it reached dynamic equilibrium
(see Figure 1). Thus, it is clear that this condition was achieved
at an earlier time for the obliquity dependent diffusion model. That
results from effective differences in diffusion times for the two models.
For the constant diffusion model, $t_d = 0.325$, which matches the nominal
(upstream)
value for the obliquity dependent diffusion model, using the
convention established in \S 2.  But, through the precursor
of that shock $\kappa$ decreases from $\kappa = 0.325$
to $\kappa = 0.1$, so that effective values of $t_d$ and $x_d$ within
the precursor and postshock flows are smaller as one proceeds downstream.
That accounts in part for the sharper and higher post shock density spike
seen in the variable diffusion model and the earlier achievement of
a steady post shock state.  Furthermore, the quicker evolution to a
CR dominated post shock state with higher compression (due to the softer
equation of state of the CR) explains why the position of the shock
is not as far advanced for the obliquity dependent diffusion model. The
shock speed decreases as the compression increases.

The more rapid shock evolution in the obliquity dependent diffusion model
is related to the point emphasized by Jokipii (1987). He argues from
test particle theory that with such
diffusion models acceleration of individual particles will be quicker
in highly oblique shocks,
raising the possibility that particles of higher energy might be
achieved in supernova remnant shocks than expected with isotropic, Bohm
diffusion. Since we have not solved the diffusion-advection equation
here we cannot comment directly on that issue except to make one point.
More rapid acceleration also leads more quickly to the development
of nonlinear feedback from the CR to the fluid. In highly oblique shocks
in low $\beta$ plasmas, that would mean a prompter onset of work done by CRs
on the tangential magnetic field.  The net result of that, of course,
is a reduction in the total work done by the flow on the CR population.
Since for $\gamma_c > \frac{4}/{3}$, $P_c$ is dominated by low energy
particles, these two behaviors are not necessarily contradictory. Clearly,
a close look at this issue is warranted, however.

\subsection{Time Dependent Effects: MHD Evolution}

In Figure 5 we present the time development of a $M_f= 10$, strong field
($\beta = 1$), CR modified oblique ($\theta_o = 60^o$) MHD shocked flow.
The evolution of the shock was
calculated with $\gamma_c = 1.41$ and nonisotropic diffusion
using $\kappa_{\parallel} = 1$ and $\kappa_{\perp} = 0.1$.
Three time frames are shown at $t/t_d = 62, 123,$ and $185$ respectively.
The plots demonstrate that the coupled CR-MHD evolution of the shock and
especially the post shock flow are complex. Basic qualitative evolution of the
fluid and the CRs are similar to CR-gasdynamical shocks, however, and
they have been discussed extensively in previously cited literature.
We concentrate here on those features that are MHD in origin.

Through the shock the tangential magnetic field is enhanced.  Mostly this is
just
compression, but in fast mode shocks there is an additional, Mach number
dependent, enhancement  through induction
generated by shear in the shock transition (\eg Kantrowitz
\& Petschek 1966).
That shear is evident in the $u_z$ jump seen through the shock.
The net tangential magnetic field change through any steady plane compression
feature is given by the simple expression
$$r_B = {B_{z2} \over B_{z1}} = r ( {M_{x}^2 - 1 \over M_{x}^2 - r}
),\eqno(3.12)$$
where $M_x$ refers to the condition upstream of the feature as measured
in the rest frame of the feature. The factor on the right in brackets
results from shear in the transition.
The shock in Figure 5 corresponds to $M_x \approx 27$, so $r_B \approx r$.
But for weaker shocks, including slow shocks the shear induced term
is more significant. Equation [3.12] also shows the well-known
difference between fast mode and slow mode waves.  Whereas the
sign of the change in tangential field and density are the same
for fast waves ($M_x^2 \ge r~>~1$), they are opposite for slow waves ($M_x^2
\le 1$).
This difference results from a reversal in the direction of the shear,
[$u_z$], relative to the normal velocity jump, [$u_x$] through the feature
(Kantrowitz \& Petschek 1966). Then the induction (see the magnetic
terms in equation [2.1]) also changes sign.  This information helps
sort out the evolution of the postshock flows seen in strong CR-MHD
oblique shocks such as that shown in Figure 5.
In this regard it is also useful to
identify from the MHD dispersion relation that through a $\beta > 1$ fast mode
compression, ($M_x$ [$u_z$] )/($M_z$ [$u_x$] ) $> 0$. The opposite
inequality applies for slow waves. For the flow described
here $M_x /M_z > 0$, so the normal velocity gradient and the shear
velocity gradient will have opposite signs in a fast compression, but
the same sign in a slow compression.

Although the shock itself in the simulations shown reaches a dynamical
steady state that is not
true for the post shock density (and magnetic field) spike.  Such spikes
result from initially strong subshocks as they adjust to new CR dominated
equilibrium conditions. At the start
this feature is a fast mode feature, since the tangential field and
gas density increase together with similar form at early times.  It
develops, after all, from the fast mode shock.  At early times one
can see that the shear velocity component, $u_z$ that leads to inductive
generation of field is negative behind the (fast mode) shock as
expected from the relation just cited.  But after the
shock has reached an equilibrium, so that the density spike is left
behind, a clear maximum in $u_z$ is present just to the left of the
peak in the density spike.  That signals the formation of a slow
wave from the density peak. This interpretation comes from the
fact that $\partial u_x / \partial x < $ across almost the entire
density spike, including the peak.  Thus it is being slowly compressed.
Hence the negative gradient in $u_z$ must represent a slow mode
compression.  This development is
even more obvious by the third and last time shown in Figure 5.
By this time, in fact, the (right facing) slow mode compression
has substantially reduced $B_z$ on its downstream side, so that
the density and magnetic spikes are clearly distinct.

Subsequently, the consequences of the slow wave are striking, in
fact, as seen in Figure 6.  This figure shows at later times the evolution
of the density spike.  One sees that the magnetic spike is eventually
confined to the forward (right) edge of the density spike.  In fact
the forward edge of the density spike has steepened into a slow mode shock.
This is demonstrated by the sharp jump in $u_x$ at this point.
The development of shocks in the transient density spike has not
happened in any of the CR-gasdynamical
shocks we have studied, so it seems to be strictly the result of the CR-MHD
nature of this calculation.  Although the density spike results from
artificial nonequilibrium initial conditions, it may still be of
much more than academic interest. Many astrophysical shocks will likely
have transient properties. We have seen such density spikes in our
simulations of SNR (Jones \& Kang 1992) and in 2D simulations of
supersonic dense clumps (Jones \& Kang 1993; Jones, Kang \& Tregillis
1993).  The velocity jump at the slow mode shock in the spike is moderately
large.  It actually does lead to a small amount of CR particle acceleration
downstream of the main shock.  That can be seen through the small "bump"
in the $P_c$ curve at the last time shown in Figure 6.
In addition, the shear and associated tangential current in the postshock
flow resembles in some
ways those inside the main shock.  There they are accompanied by drift
acceleration.  For calculations in a planar geometry this does not
appear, because the net energy flux vanishes (Jokipii 1982; Webb, Axford
\& Terasawa 1983).  However,
once that
symmetry is broken, the drifts could play important roles just as
they apparently due in perturbed oblique CR-MHD shocks (Decker 1988; Zank \etal
1990).

\section{Summary and Conclusions}

Utilizing a new numerical method built on the two-fluid model for
diffusive shock acceleration we have conducted a study of CR modified
oblique MHD plane shocks.  The results of our numerical experiments are as
follows:

\item{1.} For a fixed fast mode Mach number the particle acceleration
efficiency, $\epsilon_c$, of time asymptotic CR modified quasi perpendicular
MHD shocks decreases as the upstream obliquity, $\theta_o$, of the magnetic
field increases.  The dependence of $\epsilon_c$ on
$\theta_o$ is greater for tangential fields whose amplification through
the shock transition absorbs a significant portion of the incident
momentum flux of the CR-MHD fluid. This limits the impact of the
field on acceleration efficiency to moderately strong, quasi perpendicular
shocks in strong field (low $\beta$) plasmas.
These results confirm early analytical models and recent
numerical models of perpendicular, magnetosonic shocks.
Our experiments, however, map the relationships between
$\epsilon_c$  and the MHD and CR properties more fully than previous
studies and demonstrate that acceleration inhibition probably results
from work done on the magnetic field rather than changes in fluid
compression brought about by the presence of the field, for example.

\item{2} For quasi parallel shocks, our
results show that $\epsilon_c$ is higher at
a given fast mode Mach number when $\beta < 1$ than when $\beta > 1$.
While this would appear to suggest that strong fields enhance
acceleration efficiency in quasi parallel shocks it actually results
because in strong fields the quasi parallel fast wave speed is the
Alfv\'en speed.  The larger gas sonic Mach number is a better measure of
the strength and therefore the compression of $\beta <
1$ quasi parallel shocks.  Thus strong fields do not necessarily imply
a reduction in $\epsilon_c$. It is the tangential magnetic pressure
that is important.

\item{3.} The possibility of an anisotropic form for the CR diffusion
coefficient, $\kappa$, does not affect
the time asymptotic post shock state achieved by the CR modified MHD shocks.
However, the history of CR-MHD shocks can be considerably modified by
different forms of $\kappa$.  In particular we have examined the
effect of an obliquity dependent diffusion coefficient
$\kappa(\theta)$ on CR-MHD shock evolution.  In these models the
compression of the tangential field through the shock reduces $\kappa$
in the downstream regions and reduces the time necessary to achieve a
steady state. That is consistent with earlier suggestions that anisotropic
diffusion in quasi perpendicular shocks might allow individual cosmic
ray particles to reach higher energies in astrophysical shocks having
finite life times.  The fact that strong field, quasi perpendicular
shocks may at the same time become less efficient accelerators as the
result of nonlinear effects needs to be studied using more
complete models of diffusive acceleration.  The two outcomes
need not be contradictory, however, since the efficiency measures
total energy exchange rather than the distribution of energies.

\item{4.}  Time dependent CR-MHD effects may become significant
downstream of the shock, even once the shock itself has approached
a dynamical equilibrium.  In particular if CR modification of the
shock is sufficient to generate a postshock density spike, MHD influences
in that spike can cause substantial evolution in the spike.  We found that
a strong post shock shear develops in response to magnetic stresses in strong,
low $\beta$ oblique shocks. In more realistic
flows where planar symmetry constraints are broken this shear layer
may yield additional CR acceleration through the production of drift
currents.  Further, a slow mode shock can form on the
near edge of the post shock spike, possibly enhancing particle acceleration
from
the full structure.

\heading{Acknowledgments}

This work by TWJ and AF
was supported in part by NASA through grant NAGW -2548, the NSF
through grant AST-9100486, and by the University of Minnesota Supercomputer
Institute.
This work by DR was supported in part by NON-DIRECTED RESEARCH FUND,
Korea Research Foundation, 1993.
TWJ and AF gratefully acknowledge the hospitality of the MSI
while this work was carried out.

\vfill\eject
\heading{References}

\journal{Achterberg, A.}{1982}{A\& A}{98}{195}

\journal{Achterberg, A., Blandford, R.D, \& Periwal, V.}{1984}{A\&A}{98}{195}

\infuture{Baring, M. G., Ellison, D. C., \& Jones, F. C.}{1993}{ApJ}{in press}

\journal{Berezhko, E. G. \& Krymskii, G. F.}{1988}{Soviet
Phys.-Uspekhi}{31}{27}

\journal{Bell, A. R.}{1978}{MNRAS}{182}{147}

\journal{Blandford, R.D, \& Eichler, D.}{1987}{Phys Rept}{154}{1}

\journal{Brio, M. \& Wu, C. C.}{1988}{J. Comp. Phys.}{75}{400}

\journal{Chalov, S. V.}{1988}{Sov. Astron. Lett.}{14}{114}

\journal{Decker, R. B.}{1988}{ApJ}{324}{566}

\infuture{Donohue, D. J., Zank, G. P. \& Webb, G. M.}{1993}{ApJ}{in press}

\journal{Drury, L. O'C.}{1983}{Rep. Prog. Phys.}{46}{973}

\inbook{Drury. L. O'C.}{1987}{Proc. Sixth Int. Solar Wind Conf.}{U. J. Pizzo,
T. Holzer \& D. G. Sime}{NCAR/TN-360+Proc}{}{521}

\journal{Drury, L. O'C., \& V\"olk, H. J.}{1981}{ApJ}{248}{344}

\journal{Drury, L. O'C., \& Falle, S. A. E. G.}{1986}{MNRAS}{223}{353}

\journal{Duffy, P.}{1992}{A\& A}{262}{281}

\journal{Duffy, P., Drury, L.O'C. \& V\"olk, H. J.}{1993}{23rd ICRC,
Calgary}{vol2}{259}

\journal{Ellison, D. C. \& Eichler D.}{1984}{ApJ}{286}{691}

\journal{Falle, S. A. E. G., \& Giddings J. R.}{1987}{MNRAS}{225}{399}

\journal{Frank, A., Jones, T. W. \& Ryu, D.}{1993}{ApJS}{Dec}{???}

\journal{Harten, A.}{1983}{J. Comp. Phys.}{49}{357}

\book{Jeffrey, A.}{1966}{Magnetohydrodynamics}{London}{Oliver and Boyd}{}

\journal{Jokipii, J. R.}{1983}{ApJ}{255}{716}

\journal{Jokipii, J. R.}{1987}{ApJ}{313}{846}

\journal{Jones, F. C. \& Ellison, D. C.}{1991}{Space Sci. Rev.}{58}{259}

\inbook{Jones, T. W.}{1992}{Particle Acceleration in Cosmic Plasmas}{G. P. Zank
\& T. K. Gaisser}{American Institute of Physics}{New York}{148}

\journal{Jones, T. W.}{1993a}{ApJ}{413}{619}

\journal{Jones, T. W.}{1993b}{ApJS}{Dec}{???}

\journal{Jones, T. W. \& Kang, H.}{1990}{ApJ}{363}{499}

\journal{Jones, T. W. \& Kang, H.}{1992}{ApJ}{396}{575}

\journal{Jones, T. W. \& Kang, H.}{1993}{ApJ}{402}{560}

\infuture{Jones, T. W., Kang H. \& Tregillis, I. L.}{1993}{ApJ}{Submitted}

\infuture{Jun, B., Clarke, D. A. \& Norman, M. L.}{1993}{Preprint}{}

\journal{Kang, H.}{1993}{J.~Korea Astron.~Soc.}{26}{1}

\journal{Kang, H., \& Jones, T. W.}{1990}{ApJ}{353}{149}

\journal{Kang, H., \& Jones, T. W.}{1991}{MNRAS}{249}{439}

\journal{Kang, H., Jones, T. W., \& Ryu, D.}{1992}{ApJ}{385}{193}

\inbook{Kantrowitz, A. \& Petschek, H. E.}{1966}{Plasma Physics in Theory
and Application}{W. B. Kunkel}{McGraw Hill}{New York}{148}

\inbook{Kennel, C. F., Edmiston, J. P. \& Hada, T.}{1985}{Collisionless
Shocks in the Heliosphere: A Tutorial Review}{R. G. Stone \& B. T.
Tsurutani}{Amer. Geophys. Union}{Washington, D. C.}{1}

\journal{Kirk, J. G. \& Heavens, A. F.}{1989}{MNRAS}{239}{995}

\journal{Ostrowski, M.}{1988}{MNRAS}{233}{257}

\journal{Roe, P. L.}{1981}{J. Comp. Phys.}{43}{357}

\infuture{Ryu, D., \& Jones, T. W.}{1993}{ApJ}{to be submitted}

\journal{Ryu, D., Kang, H. \& Jones, T. W.}{1993}{ApJ}{405}{199}

\journal{Skilling, J.}{1975}{MNRAS}{172}{557}

\journal{Slavin, J. D., \& Cox, D. P.}{1992}{ApJ}{392}{131}

\journal{V\"olk, H. J., Drury, L. O'C. \& McKenzie, J. F.}{1984}{A\&
A}{130}{19}

\journal{Webb, G. M.}{1983}{A\&A}{127}{97}

\journal{Webb, G. M., Axford, W. I. \& Terasawa, T.}{1983}{ApJ}{270}{537}

\journal{Webb, G. M., Drury, L. O'C., V\"olk, H. J.}{1986}{A\& A}{160}{335}

\journal{Zank, G.P., Axford, W. I., \& McKenzie, J. F.}{1990}{A\&A}{233}{275}

\vfill\eject
\heading{Appendix: Finite Difference Algorithms}

To solve the MHD equations we have employed the
Total Variation Diminishing (TVD) method introduced by Harten (1983).
The TVD method solves the vector equation [2.1] without the
source vector ${\bmit S}$ in conservative,
finite difference form:
$$({\bmit q}_{i}^{n+1})^* = {\bmit q}_{i}^n +
{{\Delta t} \over {\Delta x}} ({\bar{\bmit{f}}}_{i+{1 \over 2}}^n({\tilde{\bmit
q}}) - {\bar{\bmit{f}}}_{i-{1 \over 2}}^n({\tilde{\bmit q}})).
\eqno({\rm A}.1)$$
As usual, $i$ is the spatial zone index; integers correspond to zone
averages (centered values).  The $n$ index refers to time step.
${\bar{\bmit{f}}}_{i\pm{1 \over 2}}^n({\tilde{\bmit q}})$ is a
4 point numerical flux function that depends on the vector ${\bmit q}$
and the original flux ${\bmit F}({\bmit q})$.
The TVD prescription of the numerical flux function
${\bar{\bmit{f}}}_{i\pm{1 \over 2}}^n({\tilde{\bmit q}})$
utilizes an approximate
MHD Riemann solver (Roe 1981; Harten 1983; Brio \& Wu 1988;
Ryu \& Jones 1993) to find
estimates of the state variables at the zone boundaries.
The method is an explicit, second order, Eulerian
finite difference scheme. The TVD
scheme also applies minimal numerical viscosity to the computed
solution resulting in strong shocks that are represented with only 2 to 3
zones. Contact and rotational discontinuities are generally
contained within $\lsim 10$ zones. For detailed descriptions
and tests for the TVD method
applied to the MHD problem see Ryu \& Jones (1993).

CR dynamical feedback on the conducting fluid is handled in a two step process
designed to preserve second order accuracy.  Gradients in $P_c$
accelerate and do work on the gas during a time step, $\Delta t$.  To
reflect that in the fluxes, ${\bar{\bmit{f}}}_{i\pm{1 \over
2}}^n({\tilde{\bmit q}})$,
and thus improve the performance of the MHD Riemann solver,
input Riemann solver momentum and energy densities are corrected to
time centered values by a simple estimate of the force and work
produced by the gradient in $P_c$,
$${\tilde{\bmit q}} = {\bmit q}^n_i +
{\Delta t\over 2}{\bmit S}^n_i. \eqno({\rm A}.2)$$
Then, after the TVD step
represented in equation [A.1] with ${\tilde{\bmit q}}$,
the state vector ${\bmit q}$ is updated
for the effects of the source terms, ${\bmit S}$; namely, $${\bmit
q}^{n+1}_{i} = ({\bmit q}^{n+1}_{i})^*+ \Delta t {\bmit
S}^n_i,\eqno({\rm A}.3)$$ where ${\bmit q}^*$ represents the solution to
equation [A.1].

The CR energy is then updated through a two step procedure, which is
also second order accurate.  In the first step we solve the Lagrangian
version of equation [2.5], $${dE_c \over dt} = {{\partial E_c} \over
{\partial t}} + u {{\partial E_c}
\over {\partial x}} = {{\partial}\over{\partial x}}\left
( \kappa{{\partial E_c}\over {\partial x}}\right )
-\gamma_c E_c{{\partial u}\over {\partial x}} + S_e, \eqno({\rm A}.4)$$
exactly as we do in the PPM two-fluid
code, using a Crank-Nicholson scheme explained in Jones \& Kang (1990).
This is followed by a "remap" step,
$$E_{c,i}^{n+1} = (E_{c,i}^{n+1})^* - u^n_i {\Delta t \over \Delta x}
(g^n_{i+{1 \over 2}}(E_c^*) - g^n_{i-{1 \over 2}}(E_c^*)),
\eqno({\rm A}.5)$$
where
$$g^n_{i\pm{1 \over 2}}({E}_c^{*}) = ({E}^n_{c,i})^{*} \pm {\Delta x \over 2}
({E}^n_{c,i})^{*'}(1 \pm {u^n_i \Delta t \over \Delta x}), \eqno({\rm A}.6)$$
$E^*_c$ is the solution to equation [A.5] and $({E}_c)^{*'}$ is the
monotone slope of ${E}_c^*$, (Van Leer 1974).
In equations [A.4 - A.6] the $x$
subscript has been dropped from $u_x$ to simplify notation.
This operator split method of solving equation [2.5] proved to
offer the best stability and accuracy among several approaches we
tried in numerical experiments during code development.

\vfill\eject
\heading{Figure Captions}

\figcap{1}{ Comparison between the analytical post-shock state (dashed
line) and the results of a time-dependent simulation with
approximately the same upstream conditions.  Both models have $M_{fs}
= 6.1$, $\theta_o = 60^o$, $\gamma_c$ = \frac{4}/{3} and $\beta =
1.0$.  The time-dependent simulation used constant $\kappa = 10$.
Shown are the density, gas pressure, magnetic pressure, normal
velocity, tangential velocity and CR pressure.  The simulation is
shown at $t/t_d = 325, 856$ , and $1377$, where the diffusion time,
$t_d$ is defined in the text.}

\bigskip

\figcap{2}{Maps of CR acceleration efficiency, $\epsilon_c$, as a function
of normal and tangential Alfv\'enic piston Mach number ($M_x$, $M_z$ )
for four different values of $\beta$.  Solid lines are contours of
$\epsilon_c$, dashed lines are contours of fast mode shock Mach number,
$M_{fs}$.  $\epsilon_c$ contour levels are marked. In all the maps
contours of $M_{fs}$ are separated by steps
of 2, with $M_{fs} = 2$ always the first contour seen in the lower
left corner of map, and $M_{fs} = 16$ in the upper right.  $\theta_{o}$
increase towards the x axis.}

\bigskip

\figcap{3}{Results from a series of $M_{fs} = 6$ runs with $\beta = 5$
(solid line) and $\beta = 1$ (dashed line) for
a range in the CR adiabatic index, $\gamma_c$.  Upper Left:
Compression ratio, $r = \rho_2/\rho_1$, as function of $\gamma_c$ for
upstream $\theta_{o} = 60^o$.  $\beta = 5$ (solid) and $\beta = 1$
(dashed).  Upper Right: CR acceleration efficiency $\epsilon_c$ as
function of $\gamma_c$ for upstream $\theta_{o} = 60^o$.  Lower Left:
CR acceleration efficiency $\epsilon_c$ as function of $\gamma_c$ for
upstream $\theta_{o} = 30^o$.  Lower Right: Normalized downstream magnetic
pressure, ${P_{B2} \over \bar P}$ as function of $\gamma_c$ for upstream
$\theta_{o} = 60^o$.  Note that the $\beta = 1$ case is always more
sensitive to $\gamma_c$ and, therefore, to compression.}

\figcap{4}{Comparison of two $\beta = 1$, $M_f = 6$, $\gamma_c = 1.41$ runs
and different models
of the CR diffusion coefficient.  Dashed line shows a constant $\kappa
=0.325$ model.  The solid line shows an obliquity
dependent model (see text) Both represent the shock structures at $t/t_d =
431$.}

\bigskip

\figcap{5}{The time evolution of a model with $M_f =10$,
$\theta_{o} = 60^o$, $\gamma_c$ = 1.41, $P_{go} = 0.032$ and $\beta = 1.0$.
$\kappa$
is obliquity dependent as in Figure 4. Structure is shown at times
$t/t_d = 62,~123$ and $185$.  Shown are the density, gas pressure,
magnetic pressure, normal velocity, tangential velocity, tangential
field, obliquity, diffusion coefficient and CR pressure.}

\bigskip

\figcap{6}{Same as Figure 5 at times $t/t_d = 308, 615$ and $923$.
 Note that here we have changed scale in $x$ to focus on the evolution
of the post shock density spike.}

\vfill\bye